\documentclass[english,reprint]{revtex4-2}
\usepackage{lmodern}
\usepackage{lmodern}
\usepackage[LGR,T1]{fontenc}
\usepackage{textcomp}
\usepackage[utf8]{inputenc}
\usepackage{babel}
\usepackage{amsmath}
\usepackage{amssymb}
\usepackage{graphicx}
\usepackage[bookmarks=false,
 breaklinks=false,pdfborder={0 0 1},backref=false,colorlinks=false]
 {hyperref}
\hypersetup{
 hidelinks,pdfcreator={LaTeX via pandoc}}

\makeatletter

\DeclareRobustCommand{\greektext}{%
  \fontencoding{LGR}\selectfont\def\encodingdefault{LGR}}
\DeclareRobustCommand{\textgreek}[1]{\leavevmode{\greektext #1}}

\@ifundefined{date}{}{\date{}}
\PassOptionsToPackage{
  pdftex,
  bookmarks=false,
  breaklinks=false,
  pdfborder={0 0 1},
  colorlinks=true,
  linkcolor=blue,
  citecolor=blue,
  urlcolor=blue
}{hyperref}

\makeatletter
\def\Dated@name{}
\makeatother
\date{}

\usepackage[cal=esstix]{mathalfa}

\makeatother

\begin{document}
\makeatletter
\let\origaddcontentsline\addcontentsline
\renewcommand{\addcontentsline}[3]{}
\makeatother
\title{From Quantum Chaos to Classical Chaos via Gain-Induced Measurement
Dynamics in a Photon Gas}
\author{V. Sharoglazova\textbf{$^{1}$\textsuperscript{}}, M. Puplauskis\textbf{$^{1}$\textsuperscript{}},
L. Hof\textbf{$^{1}$\textsuperscript{}},\textbf{\textsuperscript{}}
J. Klaers\textbf{$^{1*}$\textsuperscript{}}}
\affiliation{\textbf{\textsuperscript{1}}Adaptive Quantum Optics, MESA+ Institute
of Nanotechnology, University of Twente, Enschede, Netherlands}
\email{Corresponding author. Email: j.klaers@utwente.nl.}

\selectlanguage{english}%
\begin{abstract}
How classical chaos emerges from quantum mechanics remains a central
open question, as the unitary evolution of isolated quantum systems
forbids exponential sensitivity to initial conditions. A key insight
is that this quantum--classical link is provided by measurement processes.
In this work, we identify gain competition in a chaotic photon gas
as an operational quantum measurement that selects single motional
modes from an initial superposition through stochastic, nonlinear
amplification. We show that this mechanism naturally gives rise to
classical chaotic behavior, most notably sensitivity to initial conditions.
Our results provide a concrete physical mechanism for the quantum--classical
transition in a chaotic system and demonstrate that essential aspects
of quantum measurement---state projection, Born-rule-like selection,
and irreversibility---can naturally emerge from intrinsic gain dynamics.
\end{abstract}
\maketitle

\section{Introduction}

In classical mechanics, chaos is characterized by extreme sensitivity
to initial conditions---such as a particle's position and momentum---where
infinitesimal variations in starting parameters lead to exponentially
diverging trajectories over time. Quantum mechanics, by contrast,
treats position and momentum as complementary variables, meaning that
precise knowledge of one renders the other completely undefined. Consequently,
it is not possible to define initial conditions in the classical sense,
namely as a single point in phase space. Furthermore, the time evolution
of isolated quantum systems, as described by the Schrödinger equation,
is unitary. This means that for two quantum states $\psi_{1}(0)$
and $\psi_{2}(0)$, defining two different initial conditions, the
overlap $\left\langle \psi_{1}\left(t\right)|\psi_{2}\left(t\right)\right\rangle =\left\langle \psi_{1}\left(0\right)|U^{\dagger}\left(t\right)U\left(t\right)|\psi_{2}\left(0\right)\right\rangle =\left\langle \psi_{1}\left(0\right)|\psi_{2}\left(0\right)\right\rangle $
with $U(t)$ as the unitary time evolution operator of the system,
remains constant in time. As a result, no exponential divergence of
system states can occur. Therefore, the classical notion of sensitivity
to initial conditions has no counterpart in quantum mechanics. Instead,
quantum chaos is defined and characterized through other means, for
example, statistical properties of the energy spectrum (e.g., level
spacing distributions) and the spatial structure of wavefunctions
\cite{St90,Ha91,Sr91,St92,Al95,El95,Mo95,Am98,Fr01,Ka01,Mi01,Hu02,An06,St07,Po08,Ch09,Gu13,Ga15,Ge24}.
Thus, the notions of classical and quantum chaos differ fundamentally
in both definition and interpretation.

Still, quantum mechanics---being our most fundamental theory of nature---must
ultimately account for all phenomena observed in the classical world,
including classical chaotic motion, in accordance with the correspondence
principle. A central challenge, therefore, lies in identifying the
mechanisms by which classical chaotic behavior---such as sensitivity
to initial conditions---emerges in a quantum system. Measurement
processes are expected to play a key role in enabling such transitions
\cite{Bh00,Be01,Zu03,Ha06,Ki11}. This links the emergence of classical
chaos to one of the most fundamental open questions in quantum mechanics:
what defines, or what physical processes constitute, a measurement
\cite{Sc04,Le05,Ha22}. In our work, we experimentally demonstrate
the transition from quantum to classical chaos in the motion of particles
within a gravitational wedge system, realized with effectively two-dimensional
photons confined in a microcavity. The transition is triggered by
altering the excitation scheme of the cavity---from resonant laser
driving to non-resonant optical pumping. In the latter case, gain-induced
eigenstate selection turns infinitesimal perturbations of the pump
position into macroscopically distinct outcomes. This yields sensitivity
to initial conditions---an unambiguous indicator of classical chaos---because
the chaotic eigenmodes are mutually uncorrelated. This form of chaos
is distinct from the well-known chaotic lasing behavior in class-C
lasers, which typically concerns a single lasing mode undergoing temporal
instabilities \cite{Er10,Sc15b,Oh17}. 

\section{Gravitational-wedge for light}

\begin{figure*}
\begin{centering}
\includegraphics[width=1\textwidth]{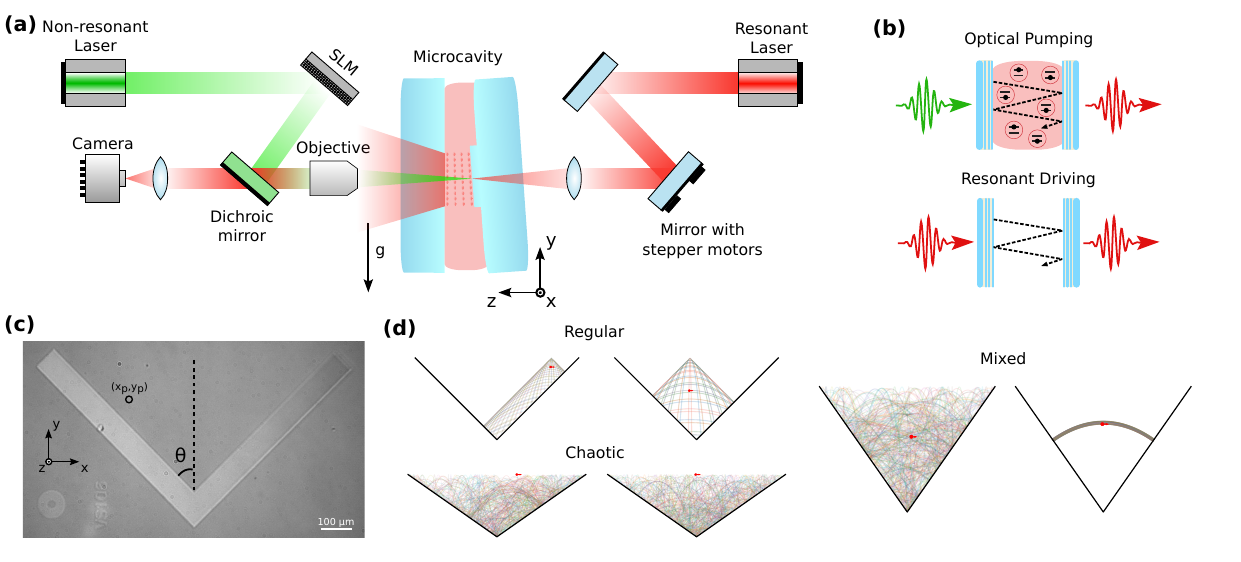}
\par\end{centering}
\caption{Gravitational wedge for two-dimensional light. (a) Schematic of the
experimental setup. The microcavity consists of two planar mirrors,
one of which is nanostructured to create an effective wedge potential
for the cavity photons. The cavity can be driven resonantly (at 655\,nm)
or optically pumped via the optical medium (at 532\,nm). A spatial
light modulator (SLM) and a motorized mirror allow for adjusting the
position of the laser spots along the cavity plane. Part of the cavity
light is transmitted through one of the mirrors and imaged onto a
camera, giving access to the system's state. One of the cavity mirrors
is tilted along the y-direction to create an effective gravitational
potential. (b) Two excitation schemes are used in the experiment.
For optical pumping (top), the optical medium is excited non-resonantly
to induce a lasing process. This process amplifies modes depending
on their overlap with the gain region (i.e., the pump spot). In the
case of resonant driving of the cavity (bottom), modes are excited
that have both spectral and spatial overlap with the driving field.
(c) Microscope image of the nanostructured mirror surface used to
create the wedge potential for the photons. The wedge is characterized
by an opening angle $\theta$. (d) Numerically calculated trajectories
of classical particles in a gravitational wedge. Depending on $\theta$,
the trajectories are regular ($\theta=45{^\circ}$) or chaotic ($\theta=55{^\circ}$).
In other cases ($\theta=35{^\circ}$), both types of motion coexist
('mixed' regime), and the behavior then depends on the initial conditions.}
\end{figure*}
Billiard systems, experimentally investigated in a variety of platforms
\cite{St90,Sr91,St92,Al95,El95,Mi01,Fr01,Ka01,Hu02,An06,Po08,Ga15,Ge24},
provide a canonical example in which chaotic dynamics arise purely
from geometric constraints. In such systems, a particle moves freely
within a bounded domain, undergoing reflections at the boundaries.
Depending on the shape of the confining region, the dynamics can be
integrable, chaotic, or---if both types of motion coexist---mixed,
in which case the behavior depends on the initial conditions. Closely
related are gravitational wedge systems, in which particles propagate
in a wedge-shaped potential landscape under the influence of an external
gravitational field \cite{Le86,Mi01}. In these systems, the opening
angle of the wedge determines whether the dynamics is integrable (45°),
chaotic (>45°), or mixed (<45°).

Our experiments are conducted with a quantum-confined photon gas in
an optical microcavity, see Fig. 1(a), which effectively realizes
quantum particles in a gravitational wedge. When the photons propagate
predominantly along the optical axis of the microcavity (paraxial
limit), projecting their motion onto the resonator plane---the x--y
plane---yields an effective description of the system as a two-dimensional,
non-relativistic gas of particles with wavevectors $(k_{x},k_{y})$
\cite{Kl10,Kl10b,Vr21,Vr21b}. In this projection, the photons acquire
an effective mass \emph{m}, which corresponds to the energy associated
with the quantized \emph{z}-component of their wavevector, $k_{z}=q\pi/D_{0}$,
where $q=1,2,3,\ldots$ is the longitudinal mode number and $D_{0}$
is the mirror separation. In our system, we find $m\approx6.9\times10^{-36}\,\text{kg}$.
Photons are introduced into the system in two distinct ways {[}Fig.
1(b){]}. The first method makes use of an active optical medium---in
our case, the dye Rhodamine 101---and non-resonant optical pumping
at a wavelength of 532\,nm, which generates electronically excited
dye molecules at a well-defined position in the cavity plane. This
leads to a lasing process in which electronic excitations are converted
into photons by stimulated emission. The process favors photons with
low transverse velocity at the pump spot, thereby maximizing the spatial
overlap of their mode functions with the gain region. In this way,
non-resonant pumping induces gain competition that typically results
in a single dominant mode. The second method involves driving the
cavity with a resonant laser beam at a wavelength of 655\,nm. Owing
to the high mode density in the cavity, this approach generally excites
multiple modes.

The transverse photon motion in the microcavity can be precisely controlled
via nanostructured mirror surfaces \cite{Ku20,Vr23}. More specifically,
a local variation $\Delta d(x,y)$ in the mirror separation generates
a corresponding potential energy $V(x,y)=-m\tilde{c}^{2}\Delta d(x,y)/D_{0}$,
with $\widetilde{c}$ as the speed of light in the optical medium.
In our experiment, we use this effect to create a wedge-shaped potential
for the photons, see Fig. 1(c). Additionally, by tilting one of the
mirrors, we introduce a linearly varying potential that effectively
acts as a gravitational potential. With \textgreek{α} as the tilting
angle of the mirror, the effective gravitational field strength is
$g_{\text{eff}}=\alpha\widetilde{c}^{2}/D_{0}\approx1.3\times10^{17}\,\textrm{ms}^{-2}$.
\begin{figure*}
\begin{centering}
\includegraphics[width=1\textwidth]{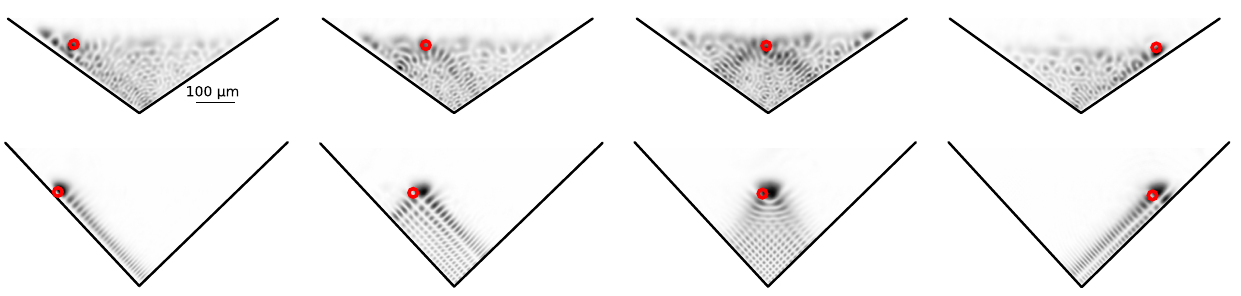}
\par\end{centering}
\caption{Regular and chaotic mode patterns. The images show experimentally
obtained mode patterns for a wedge opening angle of $\theta=55{^\circ}$
(top row) and $\theta=45{^\circ}$ (bottom row). The cavity is excited
by optical pumping, with the pump spot position indicated by the red
circle. Unlike regular patterns, chaotic patterns are ergodic, exploring
and effectively filling the entire (phase) space that is accessible
at a given energy.}
\end{figure*}

\section{Regular and chaotic motion}

Figure 2 shows experimentally obtained mode patterns for wedge potentials
with opening angles of 55° (top row) and 45° (bottom row). The system
is excited non-resonantly, and the pump-spot positions are indicated
by red circles. The opening angle of 45° gives rise to patterns characteristic
of regular (integrable) motion. In contrast, for an opening angle
of 55°, the patterns exhibit chaotic behavior, as evidenced by their
irregularity and near-ergodicity. This behavior is consistent with
what one expects for the classical counterpart of this system, see
Fig. 1(d), which shows numerically obtained trajectories of a classical
particle in different wedge geometries. 

For an opening angle of 35°, we observe both regular (Fig. 3 middle
row insets) and chaotic patterns (Fig. 3 bottom row insets), depending
on the position of the pump spot. The regular patterns correspond
to parabolic trajectories in the effective gravitational field that
reflect into themselves at the wedge boundaries. These two types of
patterns exhibit markedly different particle density distributions
(Fig. 3 middle and bottom row). The chaotic patterns follow the Porter--Thomas
distribution \cite{Al95} with good agreement. The regular patterns,
by contrast, are better described by a power-law decay.

This observation motivates a quantitative approach to distinguishing
between the two types of patterns, namely by calculating their differential
entropy, $\mathcal{S}=-\int f(\rho)\,\log f(\rho)\,d\rho$, where
\emph{f}($\rho$) denotes the probability density of observing the
particle density $\rho$ in the pattern. Note that differential entropy
\cite{Co99} is an information-theoretic quantity and does not share
the same properties as thermodynamic entropy. To enable meaningful
comparison, we normalize $\mathcal{S}$ to a dimensionless form $S\in\lbrack0,1\rbrack$
via $S=(\mathcal{S-}\mathcal{S}_{\min})/(\mathcal{S}_{\max}-\mathcal{S}_{\min})$,
where $\mathcal{S}_{\min}$ is the entropy of a delta-like distribution
(minimal disorder), and $\mathcal{S}_{\max}$ is the entropy of a
uniform distribution (maximal disorder). The resulting entropy values
are presented as a stability map in the top row of Fig. 3, which was
obtained by scanning the pump position within the wedge and computing
the differential entropy based on the experimentally observed patterns.
The map indicates that near the centre of the wedge, regular patterns
dominate, while towards the edges, chaotic patterns become prevalent.
This behavior can be understood as the result of gain competition
between the two types of motion and is discussed in more detail in
the Supplementary Material.

\section{Sensitivity to initial conditions}

We now turn to the investigation of sensitivity to initial conditions,
employing two distinct excitation schemes. While the microcavity contains
an optical medium in both cases, its role differs fundamentally between
the two. Under resonant driving, the laser couples directly to the
cavity modes, with the optical medium acting primarily as a weak absorber.
In contrast, optical pumping creates population inversion at the pump
spot, initiating a lasing process. The temporal dynamics of such processes
are well known \cite{Si86}: upon pumping, electronic excitation accumulates
in the medium, leading to spontaneous emission of photons into the
cavity modes. The buildup of electronic excitations continues until
the gain threshold is reached. At this point, the populations of the
cavity modes are exponentially amplified by stimulated emission, initiating
a phase of rapid growth. As amplification progresses, mode competition
emerges---arising from the limited availability of gain---and ultimately
results in the macroscopic occupation of a dominant cavity mode, potentially
accompanied by additional modes. For typical experimental parameters,
the amplification phase lasts of order 100\,ps \cite{Sc15a,Wa20},
after which the excess gain is depleted and the system relaxes into
a steady-state configuration, with transient dynamics largely suppressed.
This state remains stable throughout the duration of the pump pulse,
which is 26\,ns in our experiments.
\begin{figure}
\begin{raggedright}
\hspace*{-4mm}\includegraphics[width=1.1\columnwidth]{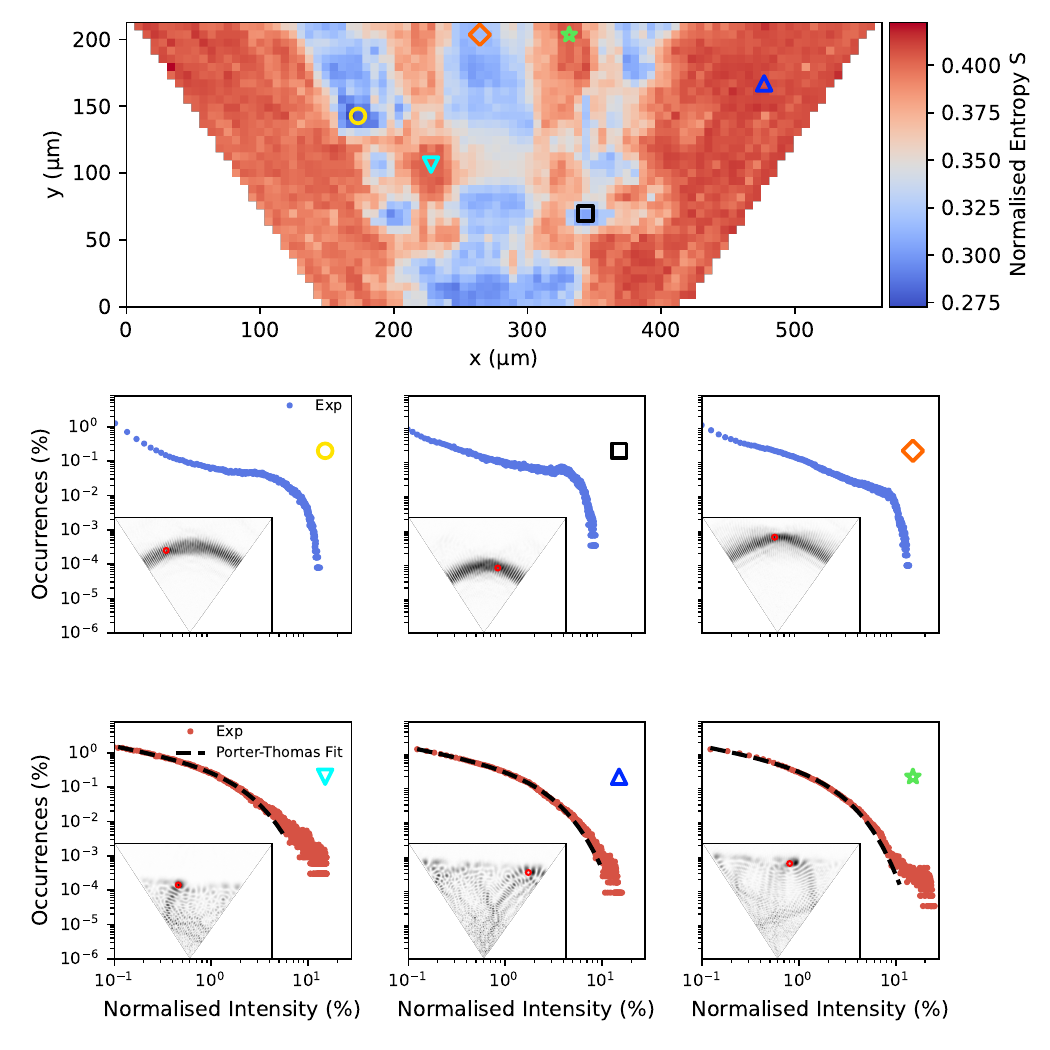}
\par\end{raggedright}
\caption{Stability map in a $\theta=35{^\circ}$ gravitational wedge. The top
row shows a map of the entropy derived from the particle density distribution
measured at different pump spot positions. High entropy values correspond
to chaotic eigenstates, while low values indicate regular ones. The
middle and bottom rows show the density distributions and camera images
for specific pump spots, indicated by distinct symbols, each corresponding
to either regular or chaotic motion. For chaotic probability density
functions, Porter--Thomas fits are shown.}
\end{figure}

For the experiments shown in Fig. 4(a), we vary the position of the
resonant driving laser along an equipotential line of the gravitational
wedge with 55° opening. To this end, we employ three different step
sizes: 11\,\textgreek{μ}m, 3.7\,\textgreek{μ}m, and 1.2\,\textgreek{μ}m.
For each step size, we compare the density patterns corresponding
to adjacent excitation spots as the laser traverses the trajectory.
The correlations between neighboring patterns are presented in Fig.
4(b-d) (red filled circles). As the step size decreases, we observe
a systematic increase in the correlations, indicating a smooth variation
of the patterns. For the smallest step size, instances of low correlation
are nearly absent providing no indication of sensitivity to initial
conditions. In contrast, the behavior under non-resonant optical pumping
is markedly different: even for the smallest step size, we frequently
observe low correlations between neighboring patterns, indicating
abrupt mode changes (Fig. 4(b-d), green filled circles). A video of
such a measurement is provided in Reference \cite{Figshare}. We interpret
this as an experimental signature of sensitivity to initial conditions,
where the latter specifically denotes the combined state of the cavity
field and the optical medium at the point when a specific photon number
is reached (see below). 

\section{Discussion}

We propose the following explanation for these results. For both excitation
schemes, the wavefunction of the particles in the cavity reaches a
steady state that can be represented as a superposition $\sum_{i}^ {}{a_{i}\left|i\right\rangle }$,
where \emph{$a_{i}$} are complex amplitudes and $\left|i\right\rangle $
denote the cavity eigenmodes. In the case of resonant laser driving,
the amplitudes \emph{a\textsubscript{i}} are determined by the spatial
and frequency overlap between the \emph{i}-th cavity mode and the
driving field. As the position of the drive is varied, these overlaps---and
hence the amplitudes---change smoothly, and no sensitivity to initial
conditions can be expected. In the case of non-resonant optical pumping,
the dynamics are driven by gain, gain saturation, and noise. Such
processes are often modeled in a semiclassical framework using stochastic
differential equations of the form \cite{Si86}:
\begin{equation}
\frac{da_{i}}{dt}=\left(\frac{g_{i}}{1+\ \sum_{j}^ {}\beta_{ij}\left|a_{j}\right|^{2}}-\kappa_{i}\right)a_{i}+\sqrt{\eta_{i}}\zeta_{i}\left(t\right)\,.\label{eq:laser}
\end{equation}
Here, \emph{$a_{i}$} is the complex mode amplitude for mode $i$,
\emph{$g_{i}$} is the gain, $\kappa_{i}$ describes losses, \emph{$\beta_{ij}$}
are the cross-gain saturation coefficients, and $\zeta_{i}(t)$ is
a complex noise term with $\left\langle \zeta_{i}\left(t\right)\zeta_{j}(t')^{*}\right\rangle =\delta_{ij}\delta(t-t')$.
The parameter \emph{$\eta_{i}$} characterizes the noise strength.
In the context of this experiment, noise primarily accounts for spontaneous
emission. Crucially, cross-gain saturation---whereby a large population
in one mode reduces the gain available to others---can give rise
to a winner-takes-all scenario in which a single eigenmode becomes
dominant. Moreover, if the exponential growth phase (before gain saturation)
is sufficiently long, the mode with the largest gain will dominate
the mode competition with certainty, despite the presence of noise.
Since the gain coefficients \emph{$g_{i}$} depend on the overlap
between the modes and the gain region, an arbitrary small change in
the pump location can steer the mode competition toward an entirely
different outcome, resulting in a discontinuous variation of the amplitudes
$a_{i}$. This by itself does not yet constitute sensitivity to initial
conditions. Such sensitivity arises because the eigenmodes $\left|i\right\rangle $
involved correspond to chaotic---and therefore largely uncorrelated---states
of motion.~Notably, this behavior stems from a deterministic mechanism,
reaffirming its consistency with the classical concept of deterministic
chaos.
\begin{figure}
\begin{centering}
\includegraphics[width=0.75\columnwidth]{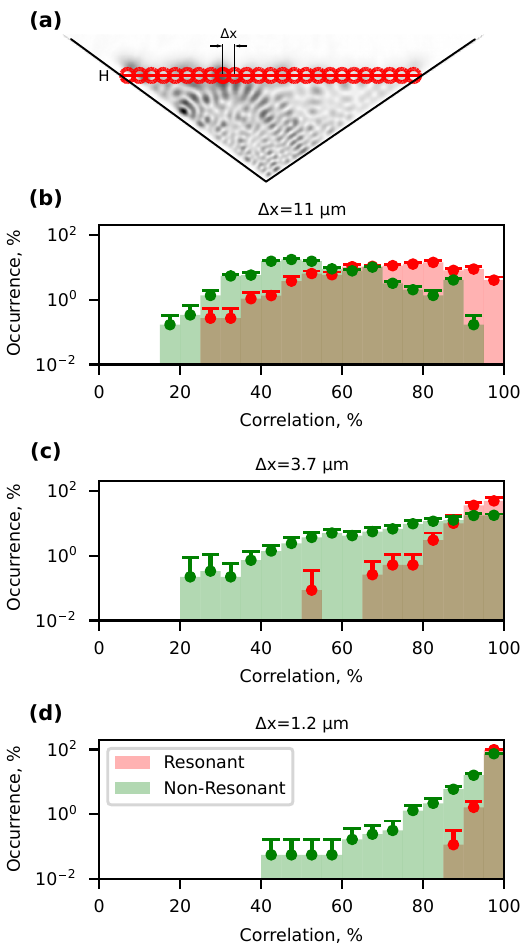}
\par\end{centering}
\caption{Sensitivity to pump spot position inside a chaotic $\theta=55{^\circ}$
wedge billiard. (a) To test for sensitivity to initial conditions,
we scan the excitation laser along an equipotential line using a fixed
step size and compare neighboring particle density patterns by calculating
their Pearson correlation coefficient (see Methods). Excitation occurs
either through resonant driving or non-resonant optical pumping. (b-d)
Correlation histograms derived from comparisons of neighboring density
patterns for three different step sizes: 11\,\textmu m, 3.7\,\textmu m,
and 1.2\,\textmu m. The histograms are normalized to relative occurrence.
With decreasing step size, correlations increase under resonant driving.
For non-resonant optical pumping, low-correlation events remain frequent,
indicating jumps between uncorrelated eigenmodes.}
\end{figure}

We now consider the experimentally demonstrated emergence of classical
chaos as an indicator of a measurement process embedded in the system's
dynamics and identify the key characteristics of this process along
with their physical origins. In particular, gain in the optical medium
selectively amplifies specific motional states, effectively converting
the initial superposition---created by spontaneous emission---into
a macroscopically populated single eigenstate. This constitutes an
effective collapse of the wavefunction and represents a defining operational
feature of quantum measurement. Since the resulting state is an energy
eigenstate, the measurement can be identified as one performed in
the energy basis. Moreover, this mode selection is, to good approximation,
consistent with the Born rule. For system dynamics governed by Eq.
\ref{eq:laser}, we define a probability \emph{$P_{i}$} for mode
\emph{$i$} to dominate the mode competition. When the modes differ
strongly in their properties, the competition typically has a trivial
winner. The more interesting case arises when the mode properties
differ only slightly, in the sense that the noise-to-gain ratios are
similar: \emph{$\eta_{i}/(g_{i}-\kappa_{i})\approx\eta_{j}/(g_{j}-\kappa_{j})$}
for all \emph{$i$} and $j$. In this regime, we define the initial
conditions of the system ($t=0$) as the combined state of the cavity
field and the optical medium at the point where the total population
across all cavity modes reaches $n_{t=0}\approx2\eta_{i}/(g_{i}-\kappa_{i})$.
In typical scenarios, this corresponds to the condition $R_{\text{stim}}\sim R_{\text{spon}}$,
where $R_{\text{stim}}$ denotes the net rate of processes proportional
to the photon number, while $R_{\text{spon}}$ accounts for spontaneous
emission. With this definition of $t=0$, the probability \emph{$P_{i}$}
that mode \emph{$i$} dominates the competition can be shown to be
well approximated by
\begin{equation}
P_{i}\approx\frac{G_{i}\left|a_{i}\left(0\right)\right|^{2}}{\sum_{j}^ {}G_{j}\left|a_{j}\left(0\right)\right|^{2}}\label{eq:Born_rule}
\end{equation}
(see Supplementary Material). Here, $|a_{i}(0)|^{2}$ denotes the
initial population of mode $i$, and the factor \emph{$G_{i}$} is
given by $G_{i}\approx\exp\left(3(g_{i}-\kappa_{i})t^{*}\right)$,
where $t^{*}$ denotes the duration of exponential growth prior to
the onset of gain saturation. This probability distribution mirrors
the Born-rule statistics in quantum mechanics, where the probability
of measuring a particular eigenstate is proportional to the squared
amplitude of its coefficient in the initial superposition. It is noted
that the factors \emph{$G_{i}$} are not identical in our experiment.
This effectively introduces a bias in the mode selection process---analogous
to a non-ideal or biased detector in a conventional measurement setup.
We further emphasize that the populations \emph{$|a_{i}(0)|^{2}$}
are not controlled in the experiment; they are randomly set by (cavity-modified)
spontaneous emission events. Furthermore, the observed measurement
process is irreversible. This irreversibility originates from the
thermodynamic asymmetry of the gain mechanism, as the amplification
of a specific mode entails heat dissipation from the gain medium to
the environment. This dissipation arises because photon emission typically
leaves the emitter molecule in a rotational--vibrational state with
excess kinetic energy relative to its equilibrium configuration, which
is eventually transferred to the environment.

The recognition that real measurements deviate from the idealized
projective postulates of textbook quantum mechanics has motivated
more general frameworks such as positive operator-valued measures
(POVMs) \cite{Bu96,Ni10}. These provide a flexible, operational description
of measurement statistics, but---like the projective formalism---they
do not explain how a single, definite outcome is realized in an individual
run. Decoherence theory complements this picture by showing how entanglement
with the environment suppresses interference and selects preferred
bases \cite{Zu03,Sc04}. While this accounts for the emergence of
classical probability distributions, it too does not address the physical
mechanism by which a single outcome is realized. This open question,
which lies at the heart of the quantum measurement problem, has motivated
the development of fundamentally different models and interpretations
of quantum mechanics \cite{Sc04,Ha22}. Experimentally, recent work
has begun to probe measurement as a physical process rather than a
mere postulate. Weak measurements extract partial information without
full collapse, enabling quantum trajectory reconstruction and revealing
the trade-off between information gain and disturbance \cite{Ah88,Lu11,Ko11}.
Quantum Zeno experiments show that frequent observations can inhibit
evolution, with the dynamics shaped by measurement strength and rate
\cite{Mi77,It90,St06,Pa15}. Real-time monitoring in platforms such
as trapped ions and superconducting qubits has revealed stochastic
quantum trajectories and quantum jumps that can even be reversed mid-flight
\cite{Mu13, Mi19}.

Our results on the quantum--classical transition in chaotic motion
demonstrate that processes with the operational signatures of a quantum
measurement can emerge from a system's intrinsic dynamics, without
invoking an explicit collapse postulate or referring to an observer---indeed,
not even a measurement device is required. In this setting, essential
features of measurement emerge, including irreversibility, Born-rule--like
statistics, and the occurrence of definite outcomes. It is this emergence
of definite outcomes that, in particular, underlies the observed sensitivity
to initial conditions in the system. The mechanism involves both nonlinearity
and noise, much like in objective-collapse models \cite{Ba13}. Unlike
in those models, however, these elements are not modifications of
quantum theory but originate from the interaction with the physical
environment. While the process studied in this work does not capture
all aspects of quantum measurement --- for instance, non-local features
--- it does demonstrate that nature provides concrete mechanisms
for measurement without recourse to an external observer.

\bibliographystyle{apsrev4-2}
\bibliography{references}

\ 

\textbf{Acknowledgments }This work received funding from the European
Research Council under the European Union's Horizon 2020 research
and innovation program (Grant Agreement No. 101001512).

\ 

\textbf{Author contributions} V.S. and M.P. performed the experiments
and analyzed the data; L.H. contributed to the project during the
initial phase; J.K. performed the theoretical modeling; V.S. and J.K.
drafted the manuscript. J.K. initiated and supervised the project.

\onecolumngrid
\newpage
\begingroup
\large
\makeatletter

\renewcommand{\uppercase}[1]{#1}

\renewcommand\section{\@startsection{section}{1}{0pt}%
  {-4.0ex \@plus -0.5ex \@minus -.2ex}
  {1.0ex \@plus 0.2ex}
  {\normalfont\large\bfseries}}

\renewcommand\subsection{\@startsection{subsection}{2}{-3.5mm}%
  {-3.0ex \@plus -0.5ex \@minus -.2ex}
  {0.8ex \@plus 0.2ex}
  {\normalfont\large\bfseries}}%

\long\def\@makecaption#1#2{%
  \vskip\abovecaptionskip
  \normalsize 
  \setlength\leftskip{0pt}%
  \setlength\rightskip{0pt}%
  \parfillskip=0pt plus 1fil\relax
  \noindent{\bfseries #1.} #2\par
  \vskip\belowcaptionskip
}

\makeatother

\ 

\vspace{10mm}

\begin{center}
{\LARGE\textbf{\textsc{Supplemental Material}}}{\LARGE\par}
\par\end{center}

\vspace{5mm}

\begin{center}
{\Large\textbf{From Quantum Chaos to Classical Chaos via Gain-Induced
Measurement Dynamics in a Photon Gas}}{\Large\par}
\par\end{center}

\begin{center}
V. Sharoglazova\textbf{$^{1}$\textsuperscript{}}, M. Puplauskis\textbf{$^{1}$\textsuperscript{}},
L. Hof\textbf{$^{1}$\textsuperscript{}},\textbf{\textsuperscript{}}
J. Klaers\textbf{$^{1*}$\textsuperscript{}}
\par\end{center}

\begin{center}
\textbf{\textsuperscript{1}}Adaptive Quantum Optics, MESA+ Institute
of Nanotechnology, University of Twente, Enschede, Netherlands
\par\end{center}

\begin{center}
$^{*}$Corresponding author. Email: j.klaers@utwente.nl.
\par\end{center}

\makeatletter
\let\addcontentsline\origaddcontentsline
\makeatother
\setcounter{page}{1}
\renewcommand{\thepage}{S\arabic{page}}
\setcounter{section}{0}
\renewcommand{\thesection}{\Roman{section}}
\renewcommand{\thesubsection}{\Roman{section}.\Alph{subsection}}
\setcounter{equation}{0}
\renewcommand{\theequation}{S\arabic{equation}}
\setcounter{figure}{0}
\renewcommand{\thefigure}{S\arabic{figure}}

\tableofcontents{}

\section{Experimental methods and data analysis}

\subsection{Experimental setup}

Our experiments are performed in an optical microcavity composed of
two distributed Bragg reflectors (DBRs), with a typical mirror separation
of approximately 16\,\textmu m. The active medium consists of Rhodamine
101 dye dissolved in ethylene glycol at a concentration of 1\,mmol/L.
One of the mirrors is mounted on three piezoelectric actuators, enabling
precise adjustment of both the cavity length and the angular alignment
between the mirrors. To measure the mutual orientation of the mirrors,
we employ additional vertical and horizontal waveguide structures.
By exciting modes in these structures and solving the corresponding
inverse problems, we can reconstruct the mirror tilts in both directions
\cite{Ma23}. The cavity length is determined by measuring the free
spectral range (FSR) of the cavity emission using a spectrometer.
For non-resonant pumping, we use a pulsed laser at 532\,nm, emitting
pulses with a full width at half maximum (FWHM) of approximately 26\,ns
at a repetition rate of 500\,Hz. A spatial light modulator (SLM)
is used to move the pump spot laterally within the cavity plane. The
camera exposure time is set to 1900\,\textmu s to ensure that each
image corresponds to a single laser pulse. The pump spot has a FWHM
of approximately 20.5\,\textmu m. A band-pass filter centered at
660\,nm with a 10\,nm bandwidth (FWHM) is placed in front of the
camera to isolate a single longitudinal mode from the cavity emission.
For resonant driving, a continuous-wave laser with a central wavelength
of 655\,nm is used. Its position in the microcavity is controlled
by a mirror mounted on stepper motors. The camera exposure time is
set to 50\,ms to achieve signal levels comparable to those obtained
under non-resonant pumping. The resonant pump spot has a FWHM of approximately
23.8\,\textmu m. All experiments are conducted under conditions that
minimize environmental disturbances. In particular, the setup is carefully
shielded from air currents to ensure maximal stability of the cavity
length.

\subsection{Sample preparation}

We use a direct laser writing nanostructuring method to shape the
surface of highly reflective mirrors at the nanometer scale \cite{Vr23}.
For this experiment, we create a wedge-shaped elevation on the mirror
surface. The height of the wedge barrier is approximately 40\,nm,
as determined via Mirau interferometry. This structure creates a reflective
potential in a microcavity setting.

\subsection{Statistical error analysis}

The stability map presented in Fig. 3 was obtained by averaging 17
independent measurements. For Fig. 4, we show results based on 15
independent measurements for non-resonant pumping and 10 for resonant
pumping. All error bars in the figures represent standard errors of
the mean.

\subsection{Differential entropy}

Each image was cropped vertically according to the pump spot height
and masked to retain only the wedge-shaped region within the potential
boundaries of the system. Images were averaged over 10 subsequent
optical pulses. To remove global brightness fluctuations, the pixel
values in each image are normalized to the average pixel value in
the picture. The pixel values are converted to a histogram, from which
we derive the probability density of the particle density $f\left(\rho\right)$.
Based on this distribution, we compute the differential Shannon entropy
\begin{equation}
\mathcal{S}=-\int f(\rho)\,\log f(\rho)\,d\rho
\end{equation}
To facilitate comparison and ensure a consistent interpretation of
entropy across different discretizations, we normalize $\mathcal{S}$
to a dimensionless measure $S\in\lbrack0,1\rbrack$ via $S=(\mathcal{S-}\mathcal{S}_{\min})/(\mathcal{S}_{\max}-\mathcal{S}_{\min})$.
The normalization relies on the limiting density of discrete points
(LDDP) framework \cite{Ja68}, which allows us to determine the appropriate
minimum and maximum entropy values, $\mathcal{S}_{\min}$ and $\mathcal{S}_{\max}$,
under the given discretization. These bounds correspond to delta-like
and uniform distributions, respectively, and account for the resolution-dependent
behavior of differential entropy estimated from discrete data.

\subsection{Mode correlations}

To quantify the correlation between mode patterns\textbf{,} we use
the Pearson correlation coefficient. For two images \emph{A}, \emph{B},
where \emph{A\textsubscript{i}}\textsubscript{,\emph{j}} denotes
the pixel value at position (\emph{i}, \emph{j}), and $\overline{A}$
is the average pixel value of image \emph{A}, the coefficient is defined
as:
\begin{equation}
r_{A,B}=\frac{\sum_{i,j}\left(A_{i,j}-\overline{A}\right)\left(B_{i,j}-\overline{B}\right)}{\sqrt{\sum_{i,j}^ {}\left(A_{i,j}-\overline{A}\right)^{2}}\sqrt{\sum_{i,j}^ {}\left(B_{i,j}-\overline{B}\right)^{2}}}\text{\ .}
\end{equation}
Here, $r_{A,B}\in\left\lbrack -1,1\right\rbrack $, where $r_{A,B}=-1$
indicates perfectly anticorrelated images, and $r_{A,B}=1$ indicates
perfectly correlated images.

\section{Born rule-like mode selection}

\subsection{Born-rule-like behavior derived from the laser equation}

Equation 1 describes the time evolution of the complex field amplitudes
\emph{a\textsubscript{i}}(\emph{t}), incorporating gain, loss, cross-gain
saturation, and noise. The general behavior in this amplification
process is as follows. Initially, noise has a significant influence
on the time evolution of the amplitudes, leading to a (partially)
non-deterministic evolution of the mode amplitudes. As the amplitudes
grow, the influence of noise diminishes, and the dynamics become largely
deterministic. In this phase, the amplitudes exhibit exponential growth.
At a certain point, this exponential growth breaks down due to cross-gain
saturation. The population in the dominant mode continues to increase
more strongly than in the inferior mode(s), leading to an effective
winner-takes-all behavior at large times.

Because of the gain-saturation term, Eq. \ref{eq:laser} does not
admit an obvious analytical solution. For identifying the dominant
mode, however, an exact solution is also not required. It suffices
to identify the dominant mode in the exponential growth phase, as
this mode will ultimately prevail once gain saturation becomes relevant.
In this regime, the gain-saturation term may be neglected, and the
equation reduces to:
\begin{align}
\frac{da(t)}{dt} & =(g-\kappa)a(t)+\sqrt{\eta}\zeta\left(t\right)\\
 & =\gamma a(t)+\sqrt{\eta}\zeta\left(t\right)
\end{align}
with 
\begin{equation}
\gamma=g-\kappa\,.
\end{equation}
We can write this equation in Itô form with a complex Wiener process
$W_{t}^{\mathbb{C}}$:
\begin{equation}
da_{t}=\gamma a_{t}dt+\sqrt{\mu}dW_{t}^{C},\ \ \ \ \ \ dW_{t}^{C}=\frac{1}{\sqrt{2}}\left(dW_{t}^{\left(1\right)}+idW_{t}^{\left(2\right)}\right),
\end{equation}
and the complex Itô rules
\begin{equation}
(dW_{t}^{C})^{2}=0,\ \ \ \ \ dW_{t}^{C}d\overline{W_{t}^{C}}=dt,\ \ \ \ \ dW_{t}^{C}dt=(dt)^{2}=0.
\end{equation}
This allows us, for instance, to evaluate the expectation value of
the intensity \emph{$|a(t)|^{2}=|a_{t}|^{2}$}. Applying the Itô calculus
to $|a_{t}|^{2}$ gives:
\begin{align}
d|a_{t}|^{2} & =\overline{a_{t}}da_{t}+a_{t}d\overline{a_{t}}+da_{t}d\overline{a_{t}}\\
 & =(2\gamma|a_{t}|^{2}+\eta)dt+\sqrt{\eta}(\overline{a_{t}}dW_{t}^{C}+a_{t}d\overline{W_{t}^{C}}).
\end{align}
Taking expectation values results in
\begin{equation}
\frac{d}{dt}\left\langle |a_{t}|^{2}\right\rangle =2\gamma\left\langle |a_{t}|^{2}\right\rangle +\eta,\label{eq:eom_photon_number}
\end{equation}
with the solution
\begin{equation}
\mu=\left\langle |a(t)|^{2}\right\rangle =|a(0)|^{2}e^{2\gamma t}+\frac{\eta}{2\gamma}(e^{2\gamma t}-1).\label{eq:mean}
\end{equation}
In a similar way, the variance of \emph{$|a_{t}|^{2}$} can be determined:
\begin{equation}
\sigma(t)^{2}=\text{Var}(|a(t)|^{2})=\frac{\eta}{\gamma}|a_{0}|^{2}e^{2\gamma t}(e^{2\gamma t}-1)+\left(\frac{\eta}{2\gamma}\right)^{2}(e^{2\gamma t}-1)^{2}.\label{eq:variance}
\end{equation}

To investigate mode competition, we now consider two modes with gains
\emph{$\gamma_{0}$} and \emph{$\gamma_{1}$} and noise strengths
$\eta_{0}$ and $\eta_{1}$. We assume that \emph{$|a_{0}(t)|^{2}$}\textsuperscript{}
and \emph{$|a_{1}(t)|^{2}$} are independent and approximately normally
distributed with means \emph{\textgreek{μ}}\textsubscript{0} and
\emph{\textgreek{μ}}\textsubscript{1} given by Eq. \ref{eq:mean}
and variances $\sigma_{0}^{2}$ and $\sigma_{1}^{2}$ given by Eq.
\ref{eq:variance}. Under this assumption, the difference between
the populations $X=X(t)=|a_{0}(t)|^{2}-|a_{1}(t)|^{2}$ itself becomes
normally distributed with mean $\mu=\mu(t)$ and standard deviation
$\sigma=\sigma(t)$. The probability that mode 0 is more strongly
populated than mode 1 can then be expressed as
\begin{align}
P_{>}\left(t\right) & =P\left(X>0\right)\\
 & =P\left(\frac{X-\mu}{\sigma}>-\frac{\mu}{\sigma}\right)\\
 & =P\left(Z>-\frac{\mu}{\sigma}\right)\\
 & =1-\Phi\left(-\frac{\mu}{\sigma}\right)\\
 & =\Phi\left(\frac{\mu}{\sigma}\right)\ ,
\end{align}
where \emph{$Z=Z(t)=(X-\mu)/\sigma$} is standard normally distributed,
and $\Phi(x)$ denotes the cumulative distribution function of the
standard normal distribution. This leads to
\begin{equation}
P_{>}\left(t\right)=\Phi\left(\frac{\langle\left|a_{0}\left(t\right)\right|^{2}\rangle-\langle\left|a_{1}\left(t\right)\right|^{2}\rangle}{\sqrt{\sigma_{0}^{2}\left(t\right)+\sigma_{1}^{2}\left(t\right)}}\right)\text{\ .}
\end{equation}

We now show that the generalized Born rule Eq. \ref{eq:Born_rule}
arises from $P_{>}(t)$ in a well-defined approximation. As an ansatz
we use
\begin{equation}
P_{B}(t)=\frac{G_{0}\left|a_{0}\left(0\right)\right|^{2}}{G_{0}\left|a_{0}\left(0\right)\right|^{2}+G_{1}\left|a_{1}\left(0\right)\right|^{2}}
\end{equation}
with $G_{i}=\exp\left(\alpha\gamma_{i}t\right)=\exp\left(\alpha(g_{i}-\kappa_{i})t\right)$
and an undetermined parameter $\alpha$. If the modes differ strongly
in their properties, the mode competition is likely to have a trivial
winner. Thus, we concentrate on a scenario in which the properties
of the modes differ only slightly in the sense that
\begin{equation}
\frac{\eta_{0}}{\gamma_{0}}\approx\frac{\eta_{1}}{\gamma_{1}}.
\end{equation}
We define $t=0$ as the moment when the total photon number reaches
a specific value of
\begin{equation}
n_{t=0}=\left|a_{0}\right|^{2}+\left|a_{1}\right|^{2}=\left(\frac{\pi}{4}+\frac{\sqrt{\pi}\sqrt{\pi+4}}{4}\right)\frac{\eta_{0}}{\gamma_{0}}\approx2\frac{\eta_{0}}{\gamma_{0}}=\frac{2\eta_{0}}{g_{0}-\kappa_{0}}.\label{eq:n_=00007Bt=00003D0=00007D}
\end{equation}
This definition, for instance when $\eta_{0}\approx\eta_{1}$, corresponds
to the condition $R_{\text{stim}}\approx2R_{\text{spon}}$, where
$R_{\text{stim}}$ denotes the net rate of processes (in both modes)
proportional to the photon number---stimulated emission and losses---while
$R_{\text{spon}}$ accounts for spontaneous emission (compare with
Eq. \ref{eq:eom_photon_number}). In our mode competition model, Eq.
\ref{eq:n_=00007Bt=00003D0=00007D} is realized by setting the initial
populations as $|a_{0}(0)|^{2}=(n_{t=0}/2)+dn$, \emph{$|a_{1}(0)|^{2}=(n_{t=0}/2)-dn$}.
With this choice, we expand $P_{>}(t)-P_{B}(t)$ as a series expansion
in \emph{$dt$} and $dn$. A straightforward but tedious calculation
shows that under these conditions
\begin{equation}
P_{>}\left(t\right)-P_{B}\left(t\right)=0+\mathcal{O}\left((dt,dn)^{3}\right),
\end{equation}
provided that
\begin{equation}
\alpha=\frac{\left(\pi+2\right)\left(\sqrt{\pi+4}-\sqrt{\pi}\right)}{\sqrt{\pi}}+\frac{\left(\sqrt{\pi+4}-\sqrt{\pi}\right)^{2}}{2}\approx3.
\end{equation}
Thus, up to terms of third order and higher, the distribution $P_{>}(t)$
coincides with the generalized Born rule Eq. \ref{eq:Born_rule}.

\begin{figure}
\begin{centering}
\includegraphics[width=0.6\textwidth]{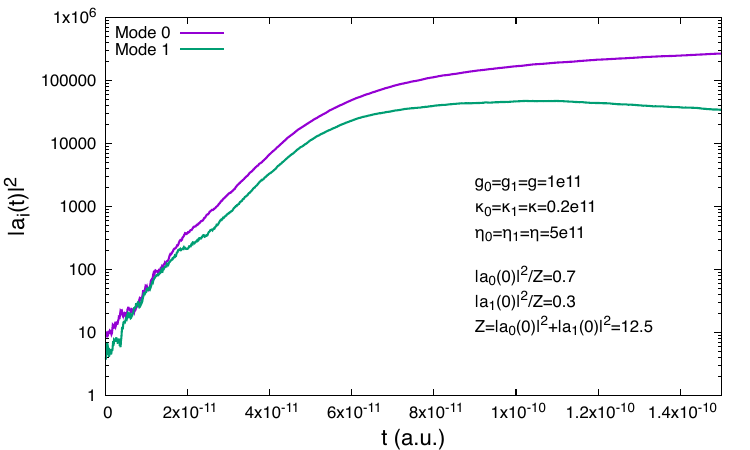}
\par\end{centering}
\caption{Stochastic trajectories for two modes following a numerical simulation
of Eq. \ref{eq:laser}.}\label{fig:2modes_trajectory}
\end{figure}

\subsection{Numerical tests}

In the following, we use numerical simulations to test this analytical
approximation. A representative example, obtained from a numerical
integration of Eq. \ref{eq:laser} for two modes, is shown in Fig.
\ref{fig:2modes_trajectory}. In this simulation (and in all subsequent
ones), the cross-gain saturation coefficients are chosen as $\beta_{ij}=1\cdot10^{-5}$
for $i=j$, and $\beta_{ij}=2\cdot10^{-5}$ for $i\neq j$. Other
parameters are given in the figure.

\begin{figure}
\begin{centering}
\includegraphics[width=0.6\textwidth]{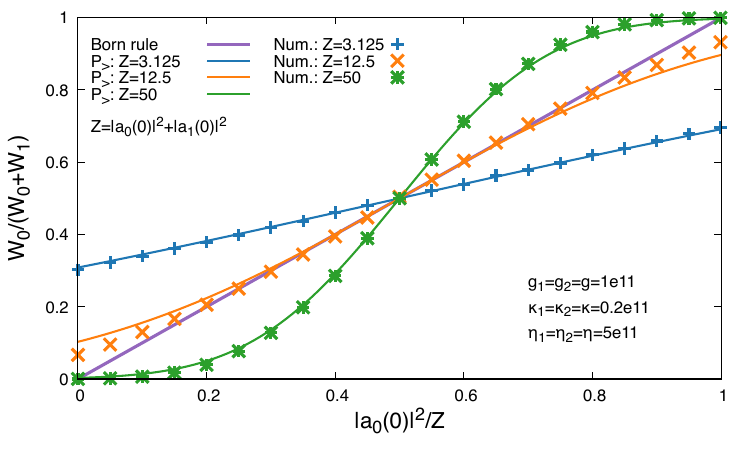}
\par\end{centering}
\caption{Relative frequency $P_{0}=W_{0}/(W_{0}+W_{1})$ with which mode 0
dominates in a two-mode competition scenario, shown as a function
of the initial relative population in mode 0 (data points). The data
is obtained from numerically integrating Eq. \ref{eq:laser}. Differently
defined initial conditions are tested, with $|a_{0}(0)|^{2}+|a_{1}(0)|^{2}=Z$
for $Z=3.125,12.5,50$. For $Z=2\eta/(g-\kappa)=12.5$, $P_{0}$ is
observed to closely match Born-rule-like behavior (a straight line
with unity slope). Solid lines show the quantity $P_{>}(t)$ with
the time \emph{$t$} set to a sufficiently large value such that $P_{>}(t)$
becomes stationary.}\label{fig:2modes}
\end{figure}
We define a relative frequency \emph{$P_{i}$} for mode \emph{i} to
win the mode competition and investigate how this quantity depends
on the system parameters. As a start, we investigate two modes with
identical gain and different initial populations. We simulate the
mode competition over a finite time interval. At the end of each simulation
run, we determine which mode has the larger amplitude \emph{$|a_{i}|$}
or intensity \emph{$|a_{i}|^{2}$}. We count the number of wins \emph{$W_{i}$}
for each mode, which yields $P_{i}=W_{i}/(W_{0}+W_{1})$ after repeating
the process many times ($N=100,000$, used in all simulations below).
Figure \ref{fig:2modes} shows \emph{$P_{i}$} as a function of the
initial relative population in mode 0 for differently defined initial
conditions with $|a_{0}|^{2}+|a_{1}|^{2}=Z$ for $Z=3.125,12.5,50$.

In these simulations, we vary the initial relative population in mode
0 while keeping \emph{$|a_{0}(0)|^{2}+|a_{1}(0)|^{2}=Z$} constant.
For $Z=50$, $P_{0}$ approaches step-like behavior, indicating that
the initially stronger mode wins with high probability. In contrast,
for $Z=3.125$, the probability distribution becomes flatter, showing
that the initial amplitude is less relevant for the outcome of the
mode competition. In an intermediate parameter regime with $Z=2\eta/(g-\kappa)=12.5$
(consistent with Eq. \ref{eq:n_=00007Bt=00003D0=00007D}), $P_{0}$
is found to be nearly identical to \emph{$|a_{0}(0)|^{2}/Z$}, corresponding
to Born rule--like behavior. The match is not perfect, as $P_{0}$
remains finite even for vanishing initial amplitude, which consequently
also leads to deviations near unit amplitude.

\begin{figure}
\begin{centering}
\includegraphics[width=0.6\textwidth]{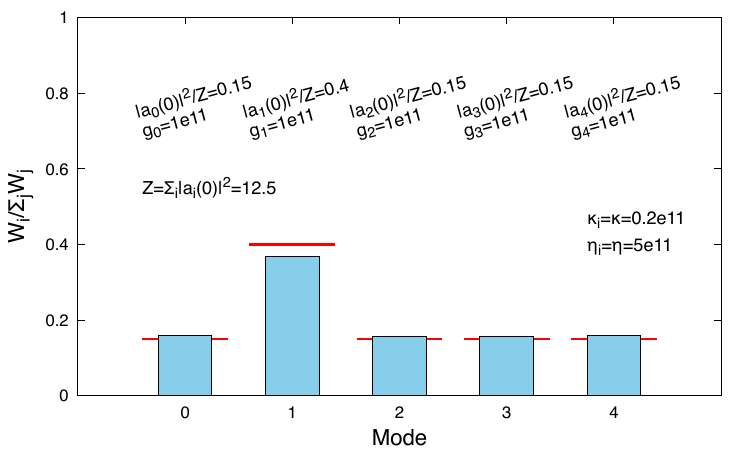}
\par\end{centering}
\caption{Relative winning frequency \emph{$P_{i}=W_{i}/\Sigma_{j}W_{j}$}
for a specific choice of initial populations \emph{$|a_{i}(t)|^{2}$}
in a 5-mode competition scenario. The red horizontal lines correspond
to Born rule-behavior.}\label{fig:5modes}
\end{figure}
The specific form of the Born rule $P_{B}(t)$ moreover suggests a
generalization for larger mode numbers, as described in Eq. \ref{eq:Born_rule}.
The validity of this generalization is tested by comparison with additional
numerical simulations. In further simulations, shown in Fig. \ref{fig:5modes},
we again use identical gain but different initial populations across
all modes and increase the number of modes to 5. Again, we observe
Born rule-like behavior in good approximation. 

\begin{figure}
\begin{centering}
\includegraphics[width=0.6\textwidth]{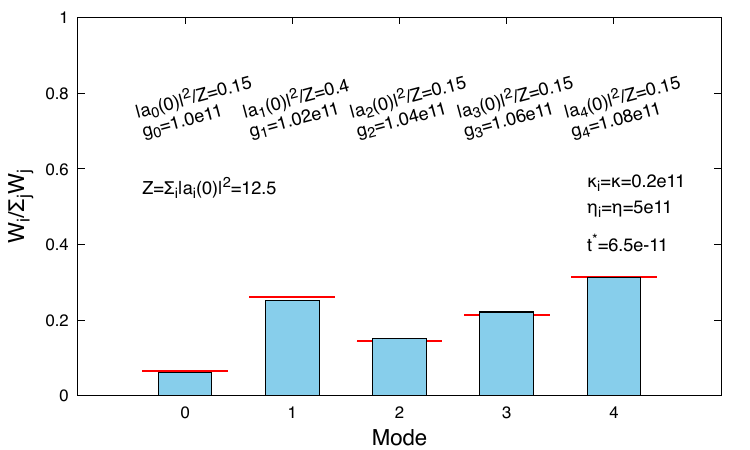}
\par\end{centering}
\caption{Relative winning frequency \emph{$P_{i}=W_{i}/\Sigma_{j}W_{j}$}
for a specific choice of gain \emph{$g_{i}$} and initial populations
$|a_{i}(t)|^{2}$ in a 5-mode competition scenario. The red horizontal
lines correspond to the generalized Born rule-behavior described by
Eq. \ref{eq:Born_rule} with $t^{*}=6.5\cdot10^{-11}$ denoting the
duration of exponential growth before gain saturation sets in (see
Fig. \ref{fig:2modes_trajectory}).}\label{fig:5modes2}
\end{figure}
Finally, we vary both the gain values and the initial populations
of the modes and determine the corresponding relative winning frequencies
(see Fig. \ref{fig:5modes2}). In this scenario, we obtain a good
match with the generalized Born rule in Eq. \ref{eq:Born_rule}. For
the time $t^{*}$, which enters Eq. \ref{eq:Born_rule} via the factors
$G_{i}=\exp\left(3(g_{i}-\kappa_{i})t^{*}\right)$, we set $t^{*}=6.5\cdot10^{-11}$.
This corresponds to the duration over which the system undergoes exponential
growth before gain saturation sets in (see also Fig. \ref{fig:2modes_trajectory},
which was simulated under the same conditions). We have explored additional
scenarios to assess the applicability of the generalized Born rule
and confirmed that it holds to a good approximation in these cases
as well. 

\section{Stability map}

In this Supplementary Information, we present a theoretical model
for the stability map shown in Fig. 3. The core idea is as follows:
placing a non-resonant pump spot at a given position $(x_{\text{pump}},y_{\text{pump}})$
within the gravitational wedge induces a competition for gain between
two types of modes---a periodic mode pattern $\psi_{p}(x,y)$ and
a chaotic mode $\psi_{c}(x,y)$ (as shown in Fig. 3). Assuming both
wave functions are normalised, the periodic mode is expected to dominate
if 
\[
|\psi_{p}(x_{\text{pump}},y_{\text{pump}})|^{2}>|\psi_{c}(x_{\text{pump}},y_{\text{pump}})|^{2},
\]
and vice versa. Our strategy is to estimate the probability densities
associated with each mode type. For the periodic modes, this is done
by analogy with the motion of classical particles in a gravitational
wedge. For the chaotic modes, we adopt a statistical description based
on known results for random interference patterns.

\subsection{Periodic orbits}

The periodic mode pattern concentrates its probability density near
a classical periodic orbit that passes through the pump spot at $(x_{\text{pump}},y_{\text{pump}})$,
see Fig. \ref{fig:per_orb}. We will construct a classical periodic
orbit that runs through a given pump spot position and estimate what
fraction of time the particle is located at the area of the pump spot.
First, we assume a periodic orbit in a constant gravitational field
of the form 
\begin{align}
y_{p}(t) & =-\frac{1}{2}gt^{2}+y_{0}\\
x(t) & =v_{x}t
\end{align}
resulting in 
\begin{equation}
y_{p}(x)=-\frac{g}{2v_{x}^{2}}x^{2}+y_{0},
\end{equation}
where $g$ is the gravitational constant, $v_{x}$ is the horizontal
velocity of the particle, and $y_{0}$ is the maximal height of the
particle. 
\begin{figure}
\begin{centering}
\includegraphics[width=0.45\textwidth]{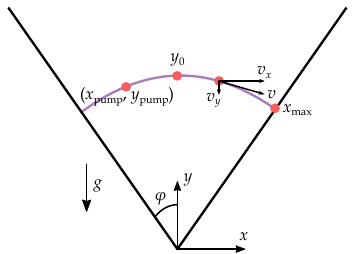} 
\par\end{centering}
\caption{Stable periodic orbit in gravitational wedge with opening angle $\varphi$.
The latter is connected to the slope of the wedge potential via $\alpha=\cot(\varphi)$.}\label{fig:per_orb}
\end{figure}

The orbit is constrained to the wedge potential described by: 
\begin{equation}
y_{w}\!\left(x\right)=\alpha|x|
\end{equation}
with $\alpha=\cot\phi$ and $\phi$ as half the opening angle of the
wedge. The position at which the particle hits the wedge potential,
corresponding to the maximal horizontal extent of the orbit, follows:
\begin{equation}
x_{\text{max}}=\frac{v_{x}}{g}\left(\sqrt{\alpha^{2}v_{x}^{2}+2gy_{0}}-\alpha v_{x}\right).
\end{equation}
At this point the trajectory needs to be orthogonal to the wedge potential:
\begin{equation}
y_{p}'(x_{\text{max}})\,y_{w}'(x_{\text{max}})=-1,
\end{equation}
where $y_{p}'=\frac{dy_{p}}{dx}$. This equation determines the horizontal
velocity along the orbit: 
\begin{equation}
v_{x}=\alpha\sqrt{\frac{2gy_{0}}{2\alpha^{2}+1}}.
\end{equation}
With this, we also determine the particle's energy: 
\begin{equation}
E=mgy_{0}+\frac{1}{2}mv_{x}^{2},
\end{equation}
and the round trip time on the orbit: 
\begin{align}
T & =4x_{\text{max}}/v_{x}\\
 & =4\sqrt{\frac{2y_{0}}{g(2\alpha^{2}+1)}}
\end{align}
For determining the time that the particle spends at the pump spot,
we need the velocity of the particle at this point. The corresponding
kinetic energy is given by: 
\begin{equation}
\frac{1}{2}mv_{\text{pump}}^{2}=E-mgy_{\text{pump}}.
\end{equation}
This gives a velocity of 
\begin{align}
v_{\text{pump}} & =\sqrt{\frac{2(E-mgy_{\text{pump}})}{m}}\\
 & =\sqrt{\frac{2g\left[(3\alpha^{2}+1)y_{0}-(2\alpha^{2}+1)y_{\text{pump}}\right]}{2\alpha^{2}+1}}
\end{align}
During one round trip the particle passes the area of the pump spot,
with spatial extent $d_{\text{pump}}$, twice. Thus, the fraction
of time that the particle spends in that region is given by: 
\begin{align}
f_{p} & =\frac{2d_{\text{pump}}/v_{\text{pump}}}{T}\\
 & =\frac{\left(2\alpha^{2}+1\right)d_{\text{pump}}}{4\sqrt{y_{0}\left[(3\alpha^{2}+1)y_{0}-(2\alpha^{2}+1)y_{\text{pump}}\right]}}\label{fraction_p}
\end{align}
Finally, the periodic orbit needs to run through the pump spot, which
leads to the following condition: 
\begin{equation}
y_{p}(x_{\text{pump}})=y_{\text{pump}}.
\end{equation}
This determines the correct value of the offset $y_{0}$: 
\begin{equation}
y_{0}=\frac{1}{2}y_{\text{pump}}+\sqrt{\frac{1}{4}y_{\text{pump}}^{2}+\left(\frac{1}{4\alpha^{2}}+\frac{1}{2}\right)x_{\text{pump}}^{2}}.\label{y0}
\end{equation}
Equations \ref{fraction_p} and \ref{y0} thus give the answer to
the question what fraction of time the particles spends at the pump
spot given a periodic orbit.

\subsection{Chaotic modes}

Unlike the periodic mode pattern, chaotic modes fill the entire volume
of the wedge. Due to random interference, they can exhibit local intensity
enhancements at specific positions---such as the location of the
pump. For any given chaotic wavefunction, the probability density
function of the intensity $I$ is 
\begin{equation}
P(I)\;=\;\frac{1}{\sqrt{2\pi\,I/\langle I\rangle}}\exp\!\Bigl(-\tfrac{I}{2\,\langle I\rangle}\Bigr).
\end{equation}
This is the well-known Porter--Thomas distribution {[}it is applicable
because, for mode patterns in a two-dimensional gravitational wedge,
the intensity does not depend on the height in the gravitational potential{]}.

With $x=I/\langle I\rangle$, the probability of finding an intensity
$x\ge\gamma$ at a given position is 
\begin{equation}
p(\gamma)=\int_{\gamma}^{\infty}\frac{1}{\sqrt{2\pi\,x}}\;e^{-x/2}\,\mathrm{d}x=1-\mathrm{erf}\!\Bigl(\sqrt{\tfrac{\gamma}{2}}\Bigr).
\end{equation}
If there are $N$ independent modes, each following the same distribution,
then the probability that \emph{no} mode exceeds $\gamma\langle I\rangle$
is 
\begin{equation}
\bigl[1-p(\gamma)\bigr]^{N}\;=\;\Bigl(\mathrm{erf}\!\bigl(\sqrt{\tfrac{\gamma}{2}}\bigr)\Bigr)^{N}.
\end{equation}
Consequently, the probability that \emph{at least one} mode exceeds
$\gamma\langle I\rangle$ is 
\begin{equation}
p(\gamma,N)=p=1-\bigl[\mathrm{erf}\!\bigl(\sqrt{\tfrac{\gamma}{2}}\bigr)\bigr]^{N}.
\end{equation}
We now solve this equation for $\gamma$. Denoting by $\mathrm{erf}^{-1}$
the inverse error function, we obtain 
\begin{equation}
\gamma=2\,\Bigl[\mathrm{erf}^{-1}\!\bigl((1-p)^{\tfrac{1}{N}}\bigr)\Bigr]^{2}.\label{gamma}
\end{equation}

\noindent Thus, with probability $p$, there exists at least one mode
(among $N$ independent chaotic modes) that reaches an intensity of
at least $\gamma\langle I\rangle$ at the pump spot location (or any
other fixed position), where $\gamma$ is given by Eq.~\ref{gamma}.

The relative time a particle spends at the location of the pump spot,
denoted by $f_{c}$, can be estimated as follows. On average, this
fraction is given by the ratio of the pump spot area to the mode volume
$V$: 
\begin{equation}
\bar{f}_{c}=\frac{\pi(d_{\text{pump}}/2)^{2}}{V},
\end{equation}
where $d_{\text{pump}}$ is the diameter of the pump spot. 
However, due to random interference effects, some modes in the ensemble
of $N$ chaotic states may exhibit significantly enhanced intensity
at the pump spot. More precisely, with probability $p$, there will
be at least one mode for which the fraction $f_{c}$ is at least 
\begin{align}
f_{c} & =\frac{\pi(d_{\text{pump}}/2)^{2}}{V}\,\gamma\\
 & =\frac{\pi(d_{\text{pump}}/2)^{2}}{y_{\text{pump}}^{2}/\alpha}\cdot2\Bigl[\mathrm{erf}^{-1}\!\bigl((1-p)^{\tfrac{1}{N}}\bigr)\Bigr]^{2},\label{fraction_c}
\end{align}
using Eq.~\ref{gamma} and $V=y_{\text{pump}}^{2}/\alpha$. To evaluate
this expression, the number of available modes $N$ still needs to
be determined.

\subsection{Density of states}

Placing a non-resonant pump spot at a specific location within the
gravitational wedge preferentially excites those modes that exhibit
a high probability density at that position. These are typically modes
whose kinetic energy is low in the vicinity of the pump. Such states
are energetically concentrated within an interval of 
\begin{equation}
\Delta E=mg\,\Delta y=mg\,d_{\text{pump}},
\end{equation}
centered around the energy 
\begin{equation}
E=mg\,y_{\text{pump}}.
\end{equation}
In the following, we estimate the number $N$ of quantum states falling
within this energy range.

To do this, we first determine the density of states of the system.
Consider a single particle of mass $m$ in two spatial dimensions,
confined by wedge-like boundaries in the $(x,y)$ plane and subject
to a linear potential $U(y)=mg\,y$. Specifically, the wedge is defined
by $y\ge\alpha|x|$. We seek the number of quantum states with energy
up to $E$, and from this, the density of states $g(E)$. In the semiclassical
approximation, the number of quantum states $N(E)$ with energy $\le E$
is given by: 
\begin{equation}
N(E)=\frac{1}{h^{2}}\int\Theta\bigl(E-H(\mathbf{q},\mathbf{p})\bigr)\,\mathrm{d}x\,\mathrm{d}y\,\mathrm{d}p_{x}\,\mathrm{d}p_{y},
\end{equation}
where $h$ is Planck's constant. At each point $(x,y)$, the kinetic
energy must satisfy 
\begin{equation}
\frac{p_{x}^{2}+p_{y}^{2}}{2m}\le E-mg\,y.
\end{equation}
Hence, 
\begin{equation}
p_{x}^{2}+p_{y}^{2}\le2m\bigl(E-mg\,y\bigr).
\end{equation}
This region in momentum space corresponds to a circle of radius 
\begin{equation}
p_{\max}(y)=\sqrt{2m\,\bigl(E-mg\,y\bigr)}.
\end{equation}
The area of this circle is 
\begin{equation}
\pi\,\bigl[p_{\max}(y)\bigr]^{2}=\pi\,\Bigl(2m\,\bigl(E-mg\,y\bigr)\Bigr).
\end{equation}
Thus, the momentum integral at $(x,y)$ yields: 
\begin{equation}
\int\mathrm{d}p_{x}\,\mathrm{d}p_{y}\,\Theta\bigl(E-H\bigr)=\pi\,\Bigl(2m\,\bigl(E-mg\,y\bigr)\Bigr).
\end{equation}
Substituting into the expression for $N(E)$, and performing the integration
over real space, we obtain: 
\begin{equation}
N(E)=\frac{1}{h^{2}}\,\int_{y\ge0}^{y\le\frac{E}{mg}}\int_{x=-y/\alpha}^{y/\alpha}\pi\,\bigl(2m\,(E-mg\,y)\bigr)\,\mathrm{d}x\,\mathrm{d}y,
\end{equation}
which evaluates to: 
\begin{equation}
N(E)=\frac{2\pi E^{3}}{3\,\alpha\,m\,g^{2}\,h^{2}}.
\end{equation}
The density of states $g(E)$ is given by the energy derivative of
$N(E)$: 
\begin{equation}
g(E)=\frac{dN}{dE}=\frac{2\pi\,E^{2}}{\alpha\,m\,g^{2}\,h^{2}}.
\end{equation}
Using this result, the number of available modes $N$ is estimated
as: 
\begin{align}
N & =g(E)\,\mathrm{d}E\\
 & =\frac{2\pi E^{2}}{\alpha mg^{2}h^{2}}\,\mathrm{d}E\\
 & =\frac{2\pi m^{2}gy_{\text{pump}}^{2}d_{\text{pump}}}{\alpha h^{2}},\label{N}
\end{align}
where we set $E=mg\,y_{\text{pump}}$ and $\mathrm{d}E=mg\,d_{\text{pump}}$.

\subsection{Numerical evaluation}

We now compare $f_{p}$ with $f_{c}$ to estimate which type of mode
dominates for a given pump spot position. For this comparison, we
use equations~\ref{fraction_p}, \ref{y0}, \ref{fraction_c}, and
\ref{N}. The parameters are set as follows: 
\begin{align*}
\alpha & =\cot\left(\frac{35}{180}\pi\right)\\
g & =1.1\cdot10^{17}\,\text{m/s}^{2}\\
m & =6.9\cdot10^{-36}\,\text{kg}\\
d_{\text{pump}} & =25\,\mu\text{m}\\
p & =0.5
\end{align*}

Based on these values, we obtain the stability map shown in Fig. \ref{fig:stability_map_theory}.

\begin{figure}
\begin{centering}
\includegraphics[width=0.7\textwidth]{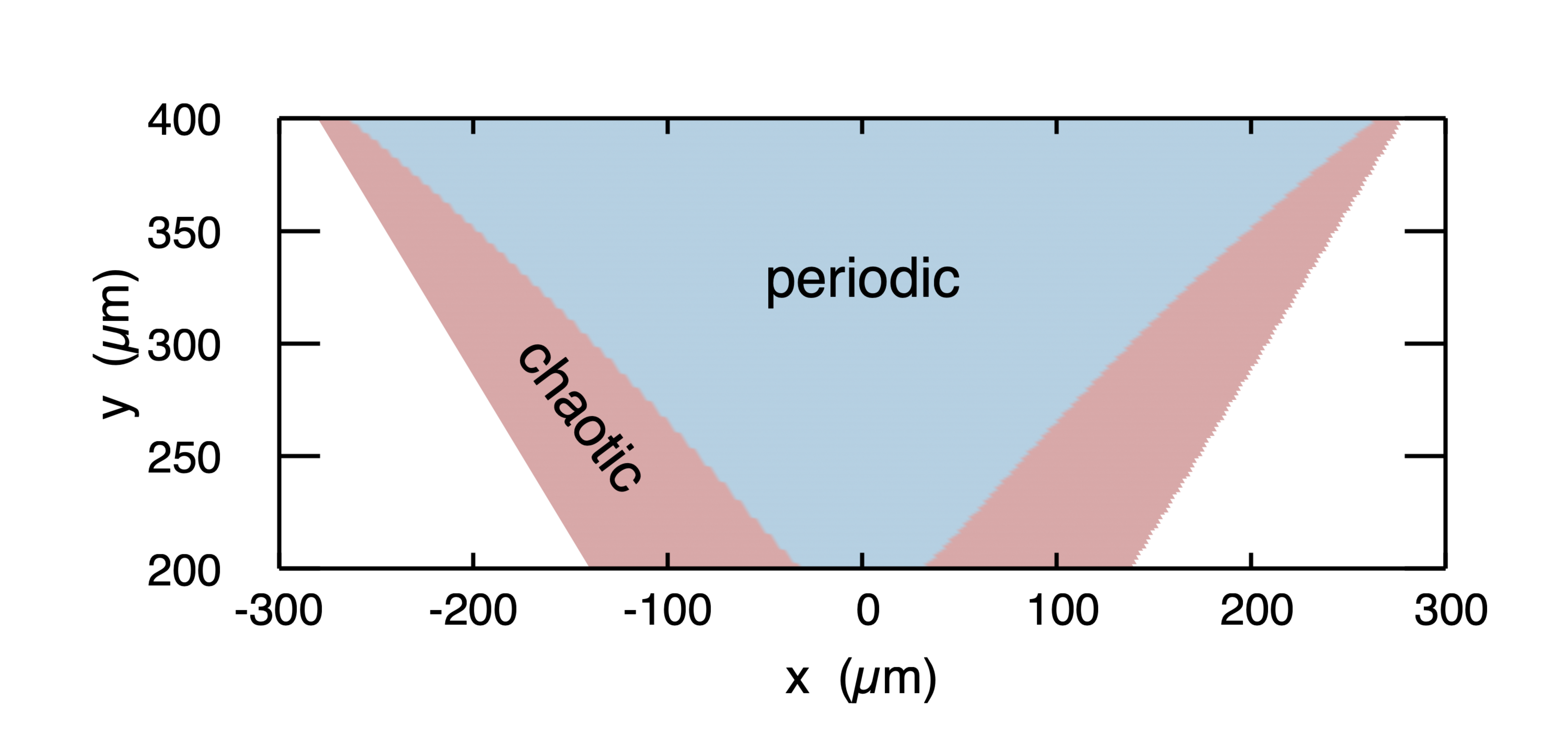} 
\par\end{centering}
\caption{Stability map. Chaotic modes dominate near the edge, whereas periodic
modes dominate closer to the center of the wedge. This is in qualitative
agreement with the experimental observations (Fig.3 in main text).}\label{fig:stability_map_theory}
\end{figure}

\endgroup

\setcounter{page}{10}
\renewcommand{\thepage}{S\arabic{page}}
\end{document}